\pdfoutput=1
\documentclass[sigconf]{acmart}
\usepackage{graphicx} 
\usepackage{todonotes}
\usepackage{comment}
\usepackage{changepage}
\usepackage{multirow}
\usepackage{adjustbox}
\usepackage{caption}
\usepackage{subcaption}
\usepackage{cleveref}
\usepackage{pbox}
\usepackage[shortlabels]{enumitem}
\long\def\comment#1{}
\setlist[enumerate]{nosep}

\newenvironment{myquote}%
  {\list{}{\leftmargin=0.05in\itshape}\item[]}%
  {\endlist}

\newcommand{\pvalue}[1]{
  \ifdim#1pt>0.05pt
    \textit{p} $={#1}$
  \else
    \ifdim#1pt<0.001pt
        \textit{\textbf{p}} $\mathbf{<.001}$
    \else
        \textit{\textbf{p}} $\mathbf{={#1}}$
    \fi
  \fi
}

\AtBeginDocument{%
  \providecommand\BibTeX{{%
    \normalfont B\kern-0.5em{\scshape i\kern-0.25em b}\kern-0.8em\TeX}}}

\copyrightyear{2023}
\acmYear{2023}
\setcopyright{acmlicensed}\acmConference[ITiCSE 2023]{Proceedings of the
2023 Conference on Innovation and Technology in Computer Science Education
V. 1}{July 8--12, 2023}{Turku, Finland}
\acmBooktitle{Proceedings of the 2023 Conference on Innovation and
Technology in Computer Science Education V. 1 (ITiCSE 2023), July 8--12,
2023, Turku, Finland}
\acmPrice{15.00}
\acmDOI{10.1145/3587102.3588842}
\acmISBN{979-8-4007-0138-2/23/07}

\begin{document}

\title{Student Usage of Q\&A Forums: Signs of Discomfort?}

\author{Naaz Sibia} 
\orcid{0000-0001-7628-7077}
\affiliation{
\institution{University of Toronto}
   \city{Toronto}
   \country{Canada}
}
\email{naaz.sibia@utoronto.ca}

\author{Angela Zavaleta Bernuy} 
\orcid{0000-0002-1228-5774}
\affiliation{
\institution{University of Toronto}
   \city{Toronto}
   \country{Canada}
}
\email{angelazb@cs.toronto.edu}

\author{Joseph Jay Williams} 
\orcid{0000-0002-9122-5242}
\affiliation{
\institution{University of Toronto}
   \city{Toronto}
   \country{Canada}
}
\email{williams@cs.toronto.edu}

\author{Michael Liut} 
\orcid{0000-0003-2965-5302}
\affiliation{
\institution{University of Toronto Mississauga}
   \city{Mississauga}
   \country{Canada}
}
\email{michael.liut@utoronto.ca}

\author{Andrew Petersen} 
\orcid{0000-0003-1337-7985}
\affiliation{
\institution{University of Toronto Mississauga}
   \city{Mississauga}
   \country{Canada}
}
\email{andrew.petersen@utoronto.ca}

\ccsdesc[500]{Social and professional topics~Computing education}
\begin{abstract}
Q\&A forums are widely used in large classes to provide scalable support. In addition to offering students a space to ask questions, these forums aim to create a community and promote engagement. 
Prior literature suggests that the way students participate in Q\&A forums varies and that most students do not actively post questions or engage in discussions. Students may display different participation behaviours depending on their comfort levels in the class. 
This paper investigates students' use of a Q\&A forum in a CS1 course. We also analyze student opinions about the forum to explain the observed behaviour, focusing on students' lack of visible participation (lurking, anonymity, private posting). We analyzed forum data collected in a CS1 course across two consecutive years and invited students to complete a survey about perspectives on their forum usage. 
Despite a small cohort of highly engaged students, we confirmed that most students do not actively read or post on the forum. We discuss students' reasons for the low level of engagement and barriers to participating visibly. Common reasons include fearing a lack of knowledge and repercussions from being visible to the student community.
\end{abstract}

\keywords{Q\&A Forums, Anonymity, Gender, Confidence}

\maketitle
\section{Introduction}
\label{intro}
Q\&A forums have become increasingly important with class sizes increasing and the shift online~\cite{barker2003online,thinnyun2021gender,joynerDoubleEdged}. They provide a scalable solution for students seeking help from course instructors and teaching assistants, and are expected to promote learning from peers (social~\cite{hill2009social} and conversation learning~\cite{boyd2013conversation} theories), and improve performance~\cite{anderson2000computer}. However, it is unclear whether participating in Q\&A forums boosts student performance or whether the correlation occurs because stronger students feel more comfortable and participate more visibly~\cite{beaudoin2002learning,mustafaraj2015visible}.

Most students do not participate actively in Q\&A forums~\cite{nandi2011active}, and this is especially true for students in large classes~\cite{ruthotto2020lurking}, from minority groups, and with low academic confidence~\cite{mazuro2011online, tafliovich2013student}. Since these forums are intended to support all students, it is important to consider what factors affect students' comfort with seeking help and if their lack of participation is influenced by factors instructors can fix.

In this work, our contributions are (1) confirmation of prior findings about forum behaviour (but in a computing context) and (2) the development of connections between student perspectives on forums and their observed behaviour. These connections lead us to (3) identify addressable factors that might increase participation and make students more comfortable. This paper differs from prior work in that it combines multiple important data sources to provide a holistic view of how students use Q\&A forums in a computing course examining participation, viewing, usage of anonymity, private posting, and general expectations. Specifically, we ask:
\begin{description}
    \item[\textbf{RQ1}:] How do students typically use (post, reply, read) a Q\&A forum in a CS1 course?
    \item[\textbf{RQ2}:] What factors are associated with a lack of participation or a lack of visible participation (anonymity, private posting, lurking)?
\end{description}

\section{Related Work}
\label{related}
Participation in Q\&A forums can encourage reflection on course material~\cite{ryle2007reflections,hill2009social}, consideration of other students' views~\cite{lake1999reducing}, and provides additional support in large courses. Student interaction with faculty~\cite{biggers2008student} and other students~\cite{barker2009exploring} have been observed as strong indicators of students' intention to major in computing, indicating the benefits of the interactions facilitated by a discussion board.

Prior work has explored levels of student participation on forums~\cite{nandi2011active, la2017understanding}, noting that most students do not actively participate or access the forum. Other studies have the impacts of both visible and invisible participation~\cite{beaudoin2002learning, mustafaraj2015visible}). Studies suggest that some students use the anonymity option on forums more frequently than others~\cite{thinnyun2021gender, joynerDoubleEdged, brigham2021gender}. However, while some groups have measured invisible participation~\cite{thinnyun2021gender, joynerDoubleEdged, beaudoin2002learning} and others have posited reasons for invisibility~\cite{dubrovsky1991equalization, karau2014understanding}, to our knowledge, no one has yet connected invisibility or a lack of participation on Q\&A forums to the specific factors in computing classrooms that theory predicts.

\subsection{Participation}
The conversation theory of learning \cite{boyd2013conversation} and social constructive learning theory stress the importance of peer-to-peer interaction in learning \cite{brown1996psychological, hill2009social}. Online Q\&A forums, in particular, have been linked lower drop-out rates and reduced feelings of isolation~\cite{lake1999reducing}, as well as better course performance ~\cite{anderson2000computer, weller2000implementing, palmer2008does}. 

Within computing, participation in course forums is linked to higher performance and retention. \citet{nandi2011active} found that high-achieving students in a CS1 course participated more on Q\&A forums than other students. Additionally, \citet{beaudoin2002learning} observed that students who made \textit{visible} contributions had higher mean grades, and \citet{mustafaraj2015visible} noticed that students who participated visibly were more likely to complete a computing MOOC. 

While the majority of engagement on forums comes from non-visible contributions or ``lurkers'' who do not actively participate in discussions~\cite{la2017understanding}, several environmental factors contribute this behaviour. These factors include low response rates, long response delays, and excessive message volume~\cite{sun2014understanding,nonnecke2001lurkers}. Lack of expertise, feelings of belonging, or academic confidence; gender; and large class sizes have also been linked to lurking~\cite{ryle2007reflections, fung2004collaborative, sun2014understanding, kuccuk2010lurking, ruthotto2020lurking,nonnecke2001lurkers}. The collective effort model~\cite{karau2014understanding} supports the idea that students avoid participating for fear that their contributions will not be perceived as unique or valuable.

\subsection{Anonymity}
Anonymous posting, another form of participation without visibility, is frequently observed on discussion forums ~\cite{thinnyun2021gender, gopal1997leveraging, joynerDoubleEdged, brigham2021gender}. The creator of Piazza (a popular Q\&A forum) added the anonymity feature based on her own experiences as a student~\cite{sobel2016class}. While this feature may help students ask questions without being embarrassed, it also hints at the status gaps or lack of academic confidence that they encounter. Work on computer-mediated communication suggests that users use anonymity to eliminate power differences and communicate without static cues, a phenomenon known as the Equalization Phenomenon~\cite{dubrovsky1991equalization}. Prior work on status theory~\cite{berger1977status} suggests that race, age, education, gender, and physical attractiveness (\textit{diffuse status characteristics}) are associated with an individual's relative status in society. Individuals with higher diffuse status are assumed to have higher competence levels, are encouraged to make more contributions in group task settings, and receive more positive reactions to their contributions ~\cite{berger1977status, carli1990gender}.
 
Evidence suggests that men~\cite{carli1990gender, flanagin2002computer,may2019gender,brooke2021trouble,terrell2017gender, mendez2018open} and people with prior programming experience feel more comfortable making contributions~\cite{tafliovich2013student, garvin2004communication}. These differences in status and encouragement to participate may incline other students to use anonymity to eliminate power differences~\cite{flanagin2002computer}. However, using anonymity can make these students disappear in these spaces.

In typical CS1 classrooms, we see an imbalance of men/women, a lack of representation, and differences in student programming expertise levels. As a result, we expect to see connections between status characteristics (e.g., gender and experience), the collective effort model, and students who are encouraged to make visible contributions. We also expect to see connections between status characteristics and support from peer networks~\cite{barker2009exploring}. 

\section{Method}
\label{dataset}
We analyzed Piazza data from two offerings of a CS1 course (first semesters of 2020 and 2021). The local review board approved the study protocol, and students provided their free and informed consent for their data to be analyzed. 

The same instructors taught both course offerings, and the course content was identical. Assessments varied slightly between terms: the same weekly homework, tests, and exam were held, but in 2020 students were required to complete 3 assignments during the semester, and in 2021, those assignments were replaced by additional exercises in weekly labs (the same concepts were tested, the workload was just distributed more evenly).

In both years, dedicated instructors and teaching assistants answered questions in Piazza, with over 90\% of questions receiving instructor responses within an average of 4 minutes. Students were asked to post any personal questions or questions that included graded code \textit{privately} to prevent cheating. When posts did not need to be private, the moderators converted them to public posts to benefit everyone.

Students in both semesters completed a survey (worth 1\% percent extra credit for their course grade) in which they were asked for their self-identified gender. In 2020, we obtained Piazza data from 454 users. 94 users identified as women, 271 identified as men, 3 identified as another gender, and 86 did not provide their gender. In 2021, we received Piazza data from 683 users. Of these users, 173 identified as women, 472 users identified as men, 7 identified as another gender, and 31 users did not provide their gender.

We collected student posts using the Piazza API and analyzed data from 4,404 questions posted in 2020 and 3,218 questions in 2021. We excluded edits to questions and answers from our analysis (as most updates to a question/answer come from the original poster), and recorded the poster's name, time of posting, post content and whether or not the user posted using the ``Anonymous to Classmates'' option (which we will call the ``anonymous'' option). For each student, we calculated the number of questions and answers posted, as well as the number of posts viewed.

\subsection{Perspectives Survey}
\label{survey}
To obtain student perspectives on the usage of anonymity and Q\&A forums in general, students in the 2021 course offering were asked to answer questions as a part of a final survey they completed. This survey was incentivized with a 1\% bonus mark in the course. 

In this survey, we asked students what platforms they preferred to use when asking questions about course material (e.g., large group chats, small friend groups, Piazza). We also asked them to describe why they use the anonymity feature, what features they pay attention to when looking at the names of posters, and what they expect from a Q\&A forum.
\subsection{General Usage Analysis}

To observe what general user participation including viewing usage (RQ1), posting choices, and upvotes, we used Python's (version 3.9.1) scikit-learn package (version 2.7) to perform agglomerative hierarchical clustering with Ward linkage and Euclidean affinity. This exploratory method minimizes variance when merging clusters. We built dendrograms of students and compared differences using feature histograms. Dendrograms visually represent the relationships between data points (or clusters) as tree-like structures. We named clusters based on their distinctive features (histograms), and drawing on related work in clusters that have been observed previously for support~\cite{la2017understanding,nandi2011active}. Although we explored smaller distance thresholds to create more clusters, we did not see prominent differences that could be labeled, but instead saw a descriptive spectrum of participation and posting.

For viewing data, we looked at the number of posts a user viewed each week. To simplify cluster identification, we grouped weeks in our data as Initial, Middle, Last, Exam, and Test weeks. These are described in Figure~\ref{fig:week-mapping}. We included both weeks surrounding the tests in each year as test weeks to account for students visiting the forums in preparation.

\begin{center}\vspace{0cm}
\includegraphics[width=227.83pt]{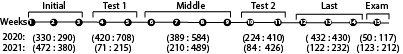}
\captionof{figure}{Timeline of weeks considered for view counts in 2020 and 2021 (public posts:private posts).}
\label{fig:week-mapping}
\end{center}\vspace{0cm}
For posting data, we looked at the number of questions a user asked publicly, anonymously, and privately. We also included the number of answers they posted publicly and anonymously. 

\subsection{Qualitative Analysis}
To supplement our quantitative data, we performed an independent qualitative analysis on open-ended survey responses to provide further insight into the relationships we observed (RQ2). We used an inductive thematic analysis approach because we were uncertain about which factors students would cite when relating their experiences on Q\&A forums. Our goal was not to create a theory, so we did not use a grounded theory analysis. Our results are discussed in Section~\ref{discussion}.

The primary analyst worked as a teaching assistant and answered Piazza questions in 2020 and 2021. The secondary analyst was a course instructor and answered questions on Piazza in 2021. Thus, we acknowledge being embedded in the environment being studied. One of us identifies as a cis-woman, and one as a cis-man. None of us live with a visible disability, one of us identifies as visibly racialized, and both of us are financially stable. Both researchers have prior experience performing qualitative work, and one of us has worked with theories of belonging and identity formation.

First, one author (primary analyst) and a research assistant tagged 10\% (123 randomly chosen) of the responses to open-ended questions, marking them as either relevant or irrelevant (blank, non-English, off-topic, or single-word answers). We calculated a Cohen's Kappa of $\kappa = 0.94$ suggesting near perfect agreement. The primary analyst coded the remaining 90\% of the responses alone. 

The primary analyst open-coded the quotes, describing and summarizing relevant information. Next, they refined the codes from the first pass and added memos. Another author (an instructor in the course), reviewed the quotes to verify the primary analyst's notations and added their own. The two authors came to a consensus on the codes. In rare cases where the authors could not reach a consensus, both views were added to the discussion.

Next, the primary analyst did a categorical coding pass, organizing and merging the initial codes into broader categories. Both authors discussed the chosen categories to arrive at a consensus. As before, if the authors could not arrive at a consensus conclusion about the category for a quote, both interpretations were added to the discussion.

Lastly, we used an individual-based sorting strategy~\cite{adu2019step} to examine what our categories represented, compared them, and categorized them further based on similarities. During this process, we had our research questions in mind. For example, the category ``fear of community response'' was placed in the ``hesitation to post'' cluster with our research question of ``why people don't post?'' in mind. We connected categories to quantitative data, existing theories, and prior work wherever it made sense after this step. We were careful to try not to justify existing theories but rather to identify theories that naturally helped to  understand our data.

\section{Results}
\subsection{Clustering}
\label{clustering}
\subsubsection{Viewing}
\label{cluster-viewing}
\begin{figure}[h]
\centering
    \begin{subfigure}[b]{0.235\textwidth}
        \centering
        \includegraphics[width=\textwidth]{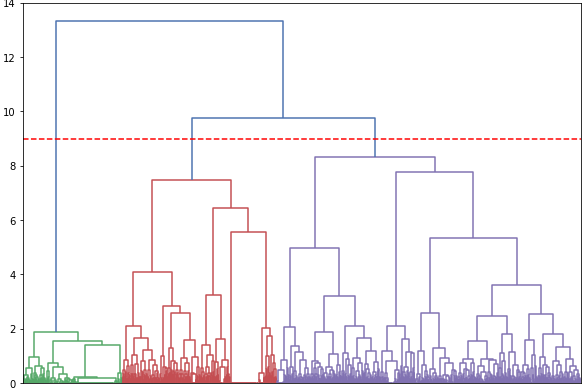}
        \caption{2021: Views over weeks.}
        \label{fig:view-dendrogram-2021}
    \end{subfigure}
    \hfill
    \begin{subfigure}[b]{0.235\textwidth}
        \centering
        \includegraphics[width=\textwidth]{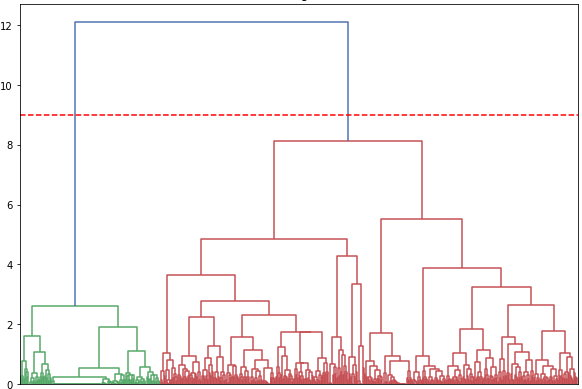}
        \caption{2020: Views over weeks.}
        \label{fig:view-dendrogram-2020}
    \end{subfigure}
    \hfill
    \begin{subfigure}[b]{0.235\textwidth}
        \centering
        \includegraphics[width=\textwidth]{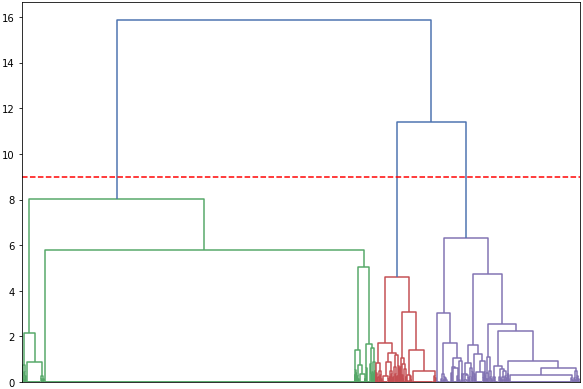}
        \caption{2021: Post visibility.}
        \label{fig:posts-dendrogram-2021}
    \end{subfigure}
    \hfill
    \begin{subfigure}[b]{0.235\textwidth}
        \centering
        \includegraphics[width=\textwidth]{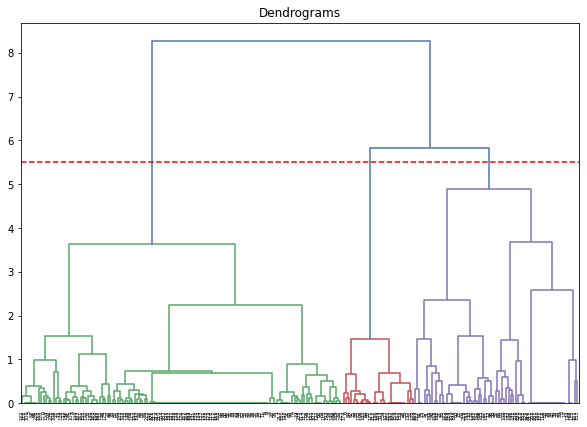}
        \caption{2020: Post visibility.}
        \label{fig:posts-dendrogram-2020}
    \end{subfigure}
    \caption{Dendrograms from cluster analysis with views and post visibility (distance thresholds indicated by the dashed red line).}
    \label{fig:dendrogram-panel}
\end{figure}

We identified three main viewing behavior clusters in the 2021 data (Figure~\ref{fig:view-dendrogram-2021}) and two in the 2020 data (Figure~\ref{fig:view-dendrogram-2020}). Cluster 1 in both years consisted of individuals who \textit{consistently viewed} posts throughout the term (more than 40 posts in the initial, middle, and last weeks), with an increase in viewing frequency before exams (around 20 posts in both years and more than 60 posts in 2021). Cluster 2 showed a \textit{decline} in views after the \textit{initial} weeks of the term (from 40 views to less than 5 views). In 2021, we discovered Cluster 3, which had spikes in views \textit{close to exams and tests} (around 3 posts each week to about 5-10 posts a day before exams and tests). We did not see a similar cluster to Cluster 3 in the 2020 data. The tests in both years were of equal weight and were conducted online. However, in 2020, students had to complete assignments every 2-3 weeks, while in 2021, they had weekly labs in addition to tests (as mentioned in Section~\ref{dataset}). As these assignments accounted for a significant portion of students' grades, they may have contributed to spikes close to assignment deadlines as well.

\subsubsection{Posting}
\label{cluster-posting}
We identified three clusters in both 2020 and 2021 (Figures~\ref{fig:posts-dendrogram-2020} and ~\ref{fig:posts-dendrogram-2021}). Cluster 1 mainly posted \textit{publicly} (63\% of their questions), while Cluster 2 mainly posted \textit{privately} (75\% and 85\% of their questions in 2020 and 2021, respectively). In Cluster 3, students posted publicly but \textit{anonymously} (74\% in 2020 and 56\% in 2021). Cluster 2 had the most private posts that did not need to be private (14\% of all their questions compared to 5\% and 6\% of Clusters 1 and 3's questions).

Cluster 1 posted the most answers (53\% in 2020 and 66\% in 2021 of total public student answers), with over 80\% of their answers being public. Cluster 2 posted 25\% of the public answers in 2021 and 45\% in 2020, with 33\% of their answers being anonymous in both years. Cluster 3 posted the fewest answers (8\% in 2021 and 2\% in 2020), with most being anonymous (over 80\%). These findings suggest that the preference for visibility level may be related to confidence in answering questions.

\subsection{Usage of Anonymity}
\label{gender-anonymity}
We conducted a Fisher Exact Two-sided test on data comparing gender and use of anonymity (Table~\ref{tab:questions-anon}). We found that women used anonymity substantially more than men when asking questions, confirming previous work~\cite{thinnyun2021gender, joynerDoubleEdged, brigham2021gender}. In both 2020 and 2021 the number of women who used anonymity all the time was about 30\% higher than men. Additionally, we observed that people either always used anonymity or never did (Figure~\ref{fig:anon-q-dis} ).
\begin{table}[ht]
    \begin{subtable}[h]{0.45\textwidth}
            \centering
        \begin{tabular}{|c|c|c|c|c|}
        \hline
        Year & Men                                                                     & Women                                                                   & p-value & Odds Ratio     \\ \hline
        2020 & \begin{tabular}[c]{@{}c@{}}N = 271 \\ (SD=0.44, \\ Mean=0.52)\end{tabular} & \begin{tabular}[c]{@{}c@{}}N = 94\\ (SD=0.33, \\ Mean=0.78)\end{tabular}  & \pvalue{.000}  & 1595  \\ \hline
        2021 & \begin{tabular}[c]{@{}c@{}}N = 472\\ (SD=0.45, \\ Mean=0.53)\end{tabular} & \begin{tabular}[c]{@{}c@{}}N = 173\\ (SD=0.32, \\ Mean=0.85)\end{tabular} & \pvalue{.000}     & 43584 \\ \hline
        \end{tabular}
    \caption{Usage of Anonymity for Questions}
    \label{tab:questions-anon}
    \end{subtable}
    \begin{subtable}[h]{0.45\textwidth}
        \begin{tabular}{|c|c|c|c|c|}
        \hline
        Year & Men                                                                     & Women                                                                  & p-value & Odds Ratio     \\ \hline
        2020 & \begin{tabular}[c]{@{}c@{}}N = 271\\ (SD=0.37, \\ Mean=0.26)\end{tabular} & \begin{tabular}[c]{@{}c@{}}N = 94\\ (SD=0.45, \\ Mean=0.50)\end{tabular} & \pvalue{.0001}    & 3.01 \\ \hline
        2021 & \begin{tabular}[c]{@{}c@{}}N = 472\\ (SD=0.40, \\ Mean=0.28)\end{tabular} & \begin{tabular}[c]{@{}c@{}}N = 173\\ (SD=0.44, \\ Mean=0.39)\end{tabular} & \pvalue{.000}   & 7.35 \\ \hline
        \end{tabular}
    \caption{Usage of Anonymity for Answers}
    \label{tab:answers-anon}
    \end{subtable}
    \caption{Usage of Anonymity - Men vs. Women}
\end{table}
\begin{center}\vspace{0cm}
\includegraphics[width=0.45\textwidth]{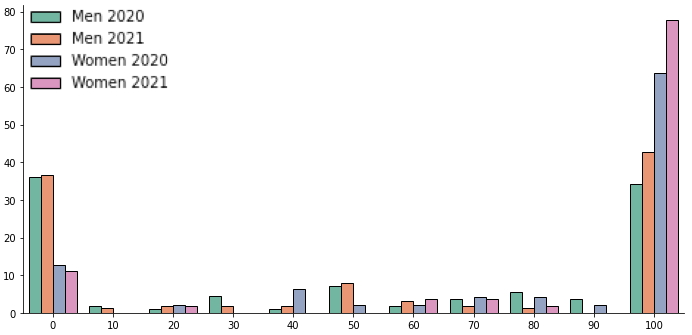}
\captionof{figure}{Proportion of the usage of "Anonymous to Classmates'' option for questions by men and women.}
\label{fig:anon-q-dis}
\end{center}\vspace{0cm}

We observed that women used anonymity for more answers than men (Table~\ref{tab:answers-anon}). The gap in anonymity usage was substantial, with around 15\% more women using anonymity for every answer than men. Interestingly, the number of students who consistently used anonymity for answering questions was around 30\% smaller than the number of students who always asked questions anonymously. This suggests that people who answer questions use anonymity less often.

\subsection{Preferred Platforms}
We surveyed students at the end of the term about their preferred platforms for asking questions about lecture. Men preferred Piazza or big public group chats, while women preferred smaller groups or other options. Statistical testing (N=368) showed a significant relationship between gender and platform choice, $\chi^2$(4)=10.6, \pvalue{.031}.

\begin{table}[ht]
\centering
\begin{adjustbox}{width=0.46\textwidth}
\begin{tabular}{|c|c|c|c|c|c|}
\hline
Option                                                    & \begin{tabular}[c]{@{}c@{}}Big unofficial \\ group chat\end{tabular} & \begin{tabular}[c]{@{}c@{}}Small group of \\ peers/friends\end{tabular} & \begin{tabular}[c]{@{}c@{}}Piazza \\ (privately)\end{tabular} & \begin{tabular}[c]{@{}c@{}}Piazza \\ (publicly)\end{tabular} & Other \\ \hline
\begin{tabular}[c]{@{}c@{}}Men\\ (N = 256)\end{tabular}   & 18\%                                                                 & 18\%                                                                    & 21\%                                                          & 27\%                                                         & 16\%  \\ \hline
\begin{tabular}[c]{@{}c@{}}Women\\ (N = 112)\end{tabular} & 8\%                                                                  & 28\%                                                                    & 19\%                                                          & 23\%                                                         & 22\%  \\ \hline
\end{tabular}
\end{adjustbox}
\caption{Platforms preferred by students to ask for help about lecture/conceptual questions}
\label{tab:help-preference}
\end{table}
 
\section{Discussion}
\label{discussion}
While some students were very active in posting and viewing posts on the Q\&A forums, as others have noted~\cite{nandi2011active, la2017understanding}, many students were not active at all (Section~\ref{clustering}). To better understand why students do not participate, we investigated their expectations of the forum, points of discomfort that discourage posting or posting visibly, and reasons for not reading forum posts. Since these forums aim to support all students, we propose reasons why students do not participate. Since students indicate using other platforms (Table ~\ref{tab:help-preference}), we also use their responses to understand which features attract them. 

\subsection{Forum Expectations}
When we asked students about their expectations of the Q\&A forum, we hoped they would cite reasons like reduced isolation, reflection on course content, asking questions, and having discussions with other students in a moderated environment. However, students want accurate answers from instructors, want to avoid being caught cheating, and focus on answers to their own questions (i.e., they do not typically read or answer other students' questions). 

Here are some quotes to illustrate these motivations. First, students indicated using the forum (Piazza) because they wanted to be confident that the answer they received is ``correct.''
\begin{myquote}
    ``I use Piazza because I am for sure going to get the correct answer within a reasonable time frame.'' - Student 9 (Man, CS Major)
\end{myquote}
Students also indicated preferring to get an answer from a TA or an Instructor because they do not trust other students to have insight.
\begin{myquote}
    ``I would usually use Piazza since the forum has profs and TAs there to help instead of other students who are also stuck.'' - Student 10 (Man, CS Major)
\end{myquote}
This suggests that students do not see other students as being able to help and may explain why students did not mention wanting to have \textit{discussions} with other students. Students prefer getting a (prompt) answer from an Instructor over working through problems with other students. This is supported by prior work, which suggests that instructor participation can decrease student discussions~\cite{aloni2018research}, and that students may want quick answers~\cite{wong2022depends}. It may also explain why many students posted privately even when they did not need to do so (Section~\ref{cluster-posting}) -- to guarantee an instructor's answer. That said, we also saw evidence that some students prefer asking questions on informal platforms where instructors are not present (Table ~\ref{tab:help-preference}), suggesting that either some students do think their peers are able to help or that their help-seeking behaviours vary based on the nature of the question. Note that our support model which is described in Section~\ref{dataset} and includes prompt responses from instructors and teaching assistants, may have influenced these behaviours.

\subsection{Barriers to Posting Publicly}
\label{discussion-posting}
From the results in Section~\ref{cluster-posting}, we know that a lot of students either did not post or predominantly posted anonymously and privately (even when unnecessary). We also see from Figure~\ref{fig:anon-q-dis} that students consistently post anonymously (or not) and from Section~\ref{cluster-posting} that students who post anonymously answer the least number of questions. Moreover, confirming work by others ~\cite{thinnyun2021gender, joynerDoubleEdged, brigham2021gender}, women post anonymously more often.

To identify some reasons behind students choosing to post anonymously, we considered the open-ended survey responses and noticed that students indicated a fear of being seen as dumb and worried that their classmates were better than them.
\begin{myquote}
    ``I feel like half of the class is better than me in regards to comprehending the material, and I do not want them to see my name when I ask a question.”  - Student 2 (Man, CS Major)
\end{myquote}
\begin{myquote}
 ``Sometimes I think when I'm behind or I'm not grasping a concept it's just me and everyone else in the class is perfectly grasping the topics at hand so I keep my questions private so only the instructors can see them" - Student 3 (Man, CS Minor)
\end{myquote}
The last quote also indicates that the board may increase a sense of social isolation rather than increasing community as intended.

We connect these quotes to the literature on Computer-Mediated Communication discussed in Section ~\ref{related}. Anonymity allows students to hide static cues and to communicate without the risks associated with displaying their identity, such as loss of social status. We also see connections to Status Theory. Students with more \textit{perceived knowledge} may make more visible contributions~\cite{tafliovich2013student}, which can further boost their community status. In Section~\ref{cluster-posting}, we see a cluster of visible students who post publicly and answer more questions, which may suggest higher confidence levels. Since seeing these visible contributions appears to make students feel less confident and drives them to post anonymously, privately, or not post at all, we argue that the status gap can widen as these students are not making visible contributions that can boost their status in the community. This suggests that we need to encourage all students to make visible contributions. Increased student participation can add stress for both instructors and students, so we do not believe that participation should be required. Instead, efforts should be made to make the environment comfortable enough for students to choose to post non-anonymously.

Students also mentioned worrying about how their peers will \textit{judge} them. They particularly appear to fear more expert students and viewed some posts as \textit{bragging}. While this is similar to the responses above, which express a fear of appearing or feeling dumb, we group these responses separately because the students are worrying more about the environment than the question being asked.
\begin{myquote}
     ``It is very likely that students who want to tell others they know more will judge me for the questions I ask and look down on me. In that situation, I'd use the anonymous feature and not be noticed by them." - Student 5 (Man, CS Major)
\end{myquote}

\begin{myquote}
    ``I've seen people laugh at questions they find ridiculous this year, and I don't think it's important why anyone should be able to identify me when asking a question that has a huge audience reach." - Student 6 (Gender and Major Unknown)
\end{myquote}

Students also mentioned that anonymity offers a way for them to ``control" the way their classmates view them. 

\begin{myquote}
    ``Even though students in the class probably won't care who is asking the questions I find it concerning that other people might view me in ways I cannot control." - Student 7 (Man, CS Major)
\end{myquote}

This is supported by work on Computer-Mediated Communication by Dubrosky et al. \cite{dubrovsky1991equalization} which suggests that people use anonymity as a form of control to hide salient characteristics.

Going back to our previous discussion about making visible contributions, these quotes offer a perspective that students may not just face confidence as a barrier to making visible contributions. They believe their activity on the forum can also negatively impact their identity in the wider community. Work by Garvin-Doxas and Barker~\cite{garvin2004communication} also highlights that students may communicate defensively and highlight having more experience/knowledge than their peers, which can negatively impact the community and students who want to ask questions.

While it is challenging to control student interactions outside the forum, this may motivate highlighting the value that a student's questions and answers add to the group (i.e., endorsements or feedback). Highlighting the importance of these contributions to the community can make students more motivated to participate~\cite{karau2014understanding}. Dividing students into exclusive spaces and groups is another possible strategy to avoid intimidation and stereotyping threat~\cite{kirkpatrick2017evaluating}. We note that this strategy successfully boosted women's confidence according to a study by Ying et al.\cite{ying2021confidence}. From Carli's work on status theory and influence \cite{carli1990gender}, communicating with people with the same perceived status can help students feel more comfortable with asking questions. This suggests that minority communities or students who lack experience may feel more comfortable communicating in exclusive spaces. Women, who show signs of discomfort through their higher usage of anonymity (Section~\ref{gender-anonymity}), also indicate preferring to seek help in small group chats (Table~\ref{tab:help-preference}), which suggests they are already seeking more exclusive spaces.

\subsection{Barriers to Reading}
Our clustering analysis in Section~\ref{cluster-viewing} reveals that many students either did not read after the initial week or read on demand (before tests and exams). From the open-ended responses on student expectations, some students mentioned that the high volume of posts, including duplicates, made it difficult to focus and read regularly. 

\begin{myquote}
    ``There are so many topics that it's hard to focus on it" - Student 13 (Man, Commerce major) 
\end{myquote}
Other students expressed frustration with having to sift through posts about others concerns; they didn't see them as potentially relevant to their own problems or learning:
\begin{myquote}
    ``I expected the environment on Piazza to be more of a general information center rather than being notified about other people's problems every so often." - Student 14 (Man, CS Major)
\end{myquote}
Addressing the issue of excessive posts is important to prevent overwhelming students and instructors, but these quotes also suggest that students may not \textit{value} other people's questions. They may view Piazza more as a \textit{help centre} for receiving individualized and tailored support from instructors and TAs, rather than a place for \textit{discussion}. This may explain why we saw clusters of on-demand students who sought help right before tests, when they had questions, in Section~\ref{cluster-viewing}. 

\subsection{Community Threats}
Tafliovich et al.~\cite{tafliovich2013student} previously discussed the stress that arises perceiving other students as more knowledgeable and experienced on Q\&A forums. Students in~\cite{tafliovich2013student} mentioned seeing people ahead of the class on Q\&A forums and seeing a small group of experts make the most visible contributions over time. We see instances of this phenomenon as students expressed frustration about others posturing on the forum:
\begin{myquote}
 ``Depending on the course, there might be students who want to tell others that they know more and ask out-of-scope questions.'' - Student 5 (Man, CS Major)
\end{myquote}

Additionally, some students spoke specifically about negative feelings associated with seeing experienced students answering questions.

\begin{myquote}
``I'm deeply insecure \& I don't want to feel like the person answering my question (experienced at coding) is doing so just to be a know-it-all. I think the endorsed badge puts an unintended negative title on the person who has it.'' - Student 16 (Man, CS Minor)
\end{myquote}

As Tafliovich et al. argue, this can become an added source of stress, which we also see is a concern for students in Section~\ref{discussion-posting}. We also note from Section~\ref{related} that visible contributions can boost a student's status in a community, and a gap in this status can lead to confidence issues that then increase this gap further as students become hesitant to make visible contributions. This means that students with lower perceived status may resort to posting privately, being anonymous, or not posting at all (clusters we observed in Section~\ref{cluster-posting}). Since these clusters, and particularly the cluster of students who post anonymously (Section~\ref{cluster-posting}), appear to answer the least number of questions, we also see hints of lower confidence.

\section{Threats to Validity}
Anonymity in this study referred to the ``Anonymous'' to classmates option on Piazza, so we cannot speak for how students feel about being visible to instructors. Furthermore, while both 2020 and 2021 had similar class structures, work was allocated slightly differently in 2020 with assignments. We also note that 2020 and 2021 were COVID-19 years (course offered completely online), which could impact results.

\section{Conclusion}
We analyzed two years of a CS1 course to identify trends in student participation and connect them to student perspectives on Q\&A forums. Using clustering techniques with quantitative Piazza data, we uncovered associations between students' posting behaviors, such as anonymity and frequency of answering questions, and their reading habits at different times during the course.
Confirming prior work~\cite{nandi2011active, la2017understanding}, most students were inactive or participated invisibly. Visible contributors answered most questions, suggesting a link to confidence. Barriers to participating visibly include threats to students' identity and low academic confidence.

Differences in participating visibly may arise from status differences~\cite{carli1990gender}, which roughly align with our findings that students feel an identity threat and do not participate if they feel they lack knowledge (status) in the community. Students with higher perceived status may make more contributions, widening the gap between visible and invisible participation. This gap can also be a source of stress for students on Q\&A forums. 

We aimed to find factors associated with a lack of visible participation that we could work on addressing in order to make Q\&A forums more comfortable. Out of the factors we found, confidence and threats to identity appear to be factors that we can strive to target with interventions. We encourage educators to explicitly work to make the online forum community less intimidating by encouraging student contributions~\cite{aloni2018research} and breaking the discussion board into smaller sub-communities~\cite{kraut2012building, shaw2013relationships, kim2013influence} where students with the same perceived status may feel more comfortable and less threatened.

\begin{acks}
This work was partially supported by the Natural Sciences and Engineering Research Council of Canada (NSERC) (\#RGPIN-2019-06968), as well as by the Office of Naval Research (ONR) (\#N00014-21-1-2576).

\end{acks}
\balance

\bibliographystyle{ACM-Reference-Format}
\bibliography{references}


\begin{thebibliography}{45}


\ifx \showCODEN    \undefined \def \showCODEN     #1{\unskip}     \fi
\ifx \showDOI      \undefined \def \showDOI       #1{#1}\fi
\ifx \showISBNx    \undefined \def \showISBNx     #1{\unskip}     \fi
\ifx \showISBNxiii \undefined \def \showISBNxiii  #1{\unskip}     \fi
\ifx \showISSN     \undefined \def \showISSN      #1{\unskip}     \fi
\ifx \showLCCN     \undefined \def \showLCCN      #1{\unskip}     \fi
\ifx \shownote     \undefined \def \shownote      #1{#1}          \fi
\ifx \showarticletitle \undefined \def \showarticletitle #1{#1}   \fi
\ifx \showURL      \undefined \def \showURL       {\relax}        \fi
\providecommand\bibfield[2]{#2}
\providecommand\bibinfo[2]{#2}
\providecommand\natexlab[1]{#1}
\providecommand\showeprint[2][]{arXiv:#2}

\bibitem[\protect\citeauthoryear{Adu}{Adu}{2019}]%
        {adu2019step}
\bibfield{author}{\bibinfo{person}{Philip Adu}.}
  \bibinfo{year}{2019}\natexlab{}.
\newblock \bibinfo{booktitle}{\emph{A Step-by-Step Guide to Qualitative Data
  Coding}}.
\newblock \bibinfo{publisher}{Routledge}.
\newblock


\bibitem[\protect\citeauthoryear{Aloni and Harrington}{Aloni and
  Harrington}{2018}]%
        {aloni2018research}
\bibfield{author}{\bibinfo{person}{Maya Aloni} {and} \bibinfo{person}{Christine
  Harrington}.} \bibinfo{year}{2018}\natexlab{}.
\newblock \showarticletitle{Research based practices for improving the
  effectiveness of asynchronous online discussion boards.}
\newblock \bibinfo{journal}{\emph{Scholarship of Teaching and Learning in
  Psychology}} \bibinfo{volume}{4}, \bibinfo{number}{4} (\bibinfo{year}{2018}),
  \bibinfo{pages}{271}.
\newblock


\bibitem[\protect\citeauthoryear{Anderson, Cheyne, Foot, Howe, Low, and
  Tolmie}{Anderson et~al\mbox{.}}{2000}]%
        {anderson2000computer}
\bibfield{author}{\bibinfo{person}{Anthony Anderson}, \bibinfo{person}{William
  Cheyne}, \bibinfo{person}{Hugh Foot}, \bibinfo{person}{Christine Howe},
  \bibinfo{person}{J Low}, {and} \bibinfo{person}{Andrew Tolmie}.}
  \bibinfo{year}{2000}\natexlab{}.
\newblock \showarticletitle{Computer support for peer-based methodology
  tutorials}.
\newblock \bibinfo{journal}{\emph{Journal of Computer Assisted Learning}}
  \bibinfo{volume}{16}, \bibinfo{number}{1} (\bibinfo{year}{2000}),
  \bibinfo{pages}{41--53}.
\newblock


\bibitem[\protect\citeauthoryear{Barker, McDowell, and Kalahar}{Barker
  et~al\mbox{.}}{2009}]%
        {barker2009exploring}
\bibfield{author}{\bibinfo{person}{Lecia~J Barker}, \bibinfo{person}{Charlie
  McDowell}, {and} \bibinfo{person}{Kimberly Kalahar}.}
  \bibinfo{year}{2009}\natexlab{}.
\newblock \showarticletitle{Exploring factors that influence computer science
  introductory course students to persist in the major}.
\newblock \bibinfo{journal}{\emph{ACM SIGCSE Bulletin}} \bibinfo{volume}{41},
  \bibinfo{number}{1} (\bibinfo{year}{2009}), \bibinfo{pages}{153--157}.
\newblock


\bibitem[\protect\citeauthoryear{Barker}{Barker}{2003}]%
        {barker2003online}
\bibfield{author}{\bibinfo{person}{Sandra Barker}.}
  \bibinfo{year}{2003}\natexlab{}.
\newblock \showarticletitle{Online discussion boards: Impacting the learning
  experience}. In \bibinfo{booktitle}{\emph{Proceedings of the fifth
  Australasian conference on Computing education-Volume 20}}. Citeseer,
  \bibinfo{pages}{53--58}.
\newblock


\bibitem[\protect\citeauthoryear{Beaudoin}{Beaudoin}{2002}]%
        {beaudoin2002learning}
\bibfield{author}{\bibinfo{person}{Michael~F Beaudoin}.}
  \bibinfo{year}{2002}\natexlab{}.
\newblock \showarticletitle{Learning or lurking?: Tracking the “invisible”
  online student}.
\newblock \bibinfo{journal}{\emph{The Internet and Higher Education}}
  \bibinfo{volume}{5}, \bibinfo{number}{2} (\bibinfo{year}{2002}),
  \bibinfo{pages}{147--155}.
\newblock


\bibitem[\protect\citeauthoryear{Berger, (Amsterdam), Fisek, and Norman}{Berger
  et~al\mbox{.}}{1977}]%
        {berger1977status}
\bibfield{author}{\bibinfo{person}{J. Berger}, \bibinfo{person}{Elsevier
  (Amsterdam)}, \bibinfo{person}{M.H. Fisek}, {and} \bibinfo{person}{R.Z.
  Norman}.} \bibinfo{year}{1977}\natexlab{}.
\newblock \bibinfo{booktitle}{\emph{{Status Characteristics and Social
  Interaction: An Expectation-states Approach}}}.
\newblock \bibinfo{publisher}{Elsevier Scientific Publishing Company}.
\newblock
\showISBNx{9780444990327}
\showLCCN{lc76025586}
\urldef\tempurl%
\url{https://books.google.ca/books?id=fi9HAAAAMAAJ}
\showURL{%
\tempurl}


\bibitem[\protect\citeauthoryear{Biggers, Brauer, and Yilmaz}{Biggers
  et~al\mbox{.}}{2008}]%
        {biggers2008student}
\bibfield{author}{\bibinfo{person}{Maureen Biggers}, \bibinfo{person}{Anne
  Brauer}, {and} \bibinfo{person}{Tuba Yilmaz}.}
  \bibinfo{year}{2008}\natexlab{}.
\newblock \showarticletitle{Student Perceptions of Computer Science: A
  Retention Study Comparing Graduating Seniors vs. CS Leavers}.
\newblock \bibinfo{journal}{\emph{ACM SIGCSE bulletin}} \bibinfo{volume}{40},
  \bibinfo{number}{1} (\bibinfo{year}{2008}), \bibinfo{pages}{402--406}.
\newblock


\bibitem[\protect\citeauthoryear{Boyd}{Boyd}{2013}]%
        {boyd2013conversation}
\bibfield{author}{\bibinfo{person}{Gary~McIntyre Boyd}.}
  \bibinfo{year}{2013}\natexlab{}.
\newblock \showarticletitle{Conversation theory}.
\newblock In \bibinfo{booktitle}{\emph{Handbook of research on educational
  communications and technology}}. \bibinfo{publisher}{Routledge},
  \bibinfo{pages}{189--207}.
\newblock


\bibitem[\protect\citeauthoryear{Brigham and Porquet-Lupine}{Brigham and
  Porquet-Lupine}{2021}]%
        {brigham2021gender}
\bibfield{author}{\bibinfo{person}{Madison Brigham} {and}
  \bibinfo{person}{Jo{\"e}l Porquet-Lupine}.} \bibinfo{year}{2021}\natexlab{}.
\newblock \showarticletitle{{Gender Differences in Class Participation in Core
  CS Courses}}. In \bibinfo{booktitle}{\emph{Proceedings of the 26th ACM
  Conference on Innovation and Technology in Computer Science Education V. 1}}.
  \bibinfo{pages}{478--483.}
\newblock


\bibitem[\protect\citeauthoryear{Brooke}{Brooke}{2021}]%
        {brooke2021trouble}
\bibfield{author}{\bibinfo{person}{SJ Brooke}.}
  \bibinfo{year}{2021}\natexlab{}.
\newblock \showarticletitle{{Trouble in programmer’s paradise: gender-biases
  in sharing and recognising technical knowledge on Stack Overflow}}.
\newblock \bibinfo{journal}{\emph{Information, Communication \& Society}}
  \bibinfo{volume}{24}, \bibinfo{number}{14} (\bibinfo{year}{2021}),
  \bibinfo{pages}{2091--2112}.
\newblock


\bibitem[\protect\citeauthoryear{Brown and Campione}{Brown and
  Campione}{1996}]%
        {brown1996psychological}
\bibfield{author}{\bibinfo{person}{Ann~L Brown} {and} \bibinfo{person}{Joseph~C
  Campione}.} \bibinfo{year}{1996}\natexlab{}.
\newblock \bibinfo{booktitle}{\emph{Psychological theory and the design of
  innovative learning environments: On procedures, principles, and systems.}}
\newblock \bibinfo{publisher}{Lawrence Erlbaum Associates, Inc}.
\newblock


\bibitem[\protect\citeauthoryear{Carli}{Carli}{1990}]%
        {carli1990gender}
\bibfield{author}{\bibinfo{person}{Linda~L Carli}.}
  \bibinfo{year}{1990}\natexlab{}.
\newblock \showarticletitle{{Gender, language, and influence.}}
\newblock \bibinfo{journal}{\emph{Journal of Personality and Social
  Psychology}} \bibinfo{volume}{59}, \bibinfo{number}{5}
  (\bibinfo{year}{1990}), \bibinfo{pages}{941}.
\newblock


\bibitem[\protect\citeauthoryear{Dubrovsky, Kiesler, and Sethna}{Dubrovsky
  et~al\mbox{.}}{1991}]%
        {dubrovsky1991equalization}
\bibfield{author}{\bibinfo{person}{Vitaly~J Dubrovsky}, \bibinfo{person}{Sara
  Kiesler}, {and} \bibinfo{person}{Beheruz~N Sethna}.}
  \bibinfo{year}{1991}\natexlab{}.
\newblock \showarticletitle{{The Equalization Phenomenon: Status Effects in
  Computer-Mediated and Face-to-Face Decision-Making Groups}}.
\newblock \bibinfo{journal}{\emph{Human-computer interaction}}
  \bibinfo{volume}{6}, \bibinfo{number}{2} (\bibinfo{year}{1991}),
  \bibinfo{pages}{119--146}.
\newblock


\bibitem[\protect\citeauthoryear{Flanagin, Tiyaamornwong, O'Connor, and
  Seibold}{Flanagin et~al\mbox{.}}{2002}]%
        {flanagin2002computer}
\bibfield{author}{\bibinfo{person}{Andrew~J Flanagin}, \bibinfo{person}{Vanessa
  Tiyaamornwong}, \bibinfo{person}{Joan O'Connor}, {and}
  \bibinfo{person}{David~R Seibold}.} \bibinfo{year}{2002}\natexlab{}.
\newblock \showarticletitle{{Computer-Mediated Group Work: The Interaction of
  Sex and Anonymity}}.
\newblock \bibinfo{journal}{\emph{Communication Research}}
  \bibinfo{volume}{29}, \bibinfo{number}{1} (\bibinfo{year}{2002}),
  \bibinfo{pages}{66--93}.
\newblock


\bibitem[\protect\citeauthoryear{Fung*}{Fung*}{2004}]%
        {fung2004collaborative}
\bibfield{author}{\bibinfo{person}{Yvonne~YH Fung*}.}
  \bibinfo{year}{2004}\natexlab{}.
\newblock \showarticletitle{Collaborative online learning: Interaction patterns
  and limiting factors}.
\newblock \bibinfo{journal}{\emph{Open Learning: The Journal of Open, Distance
  and e-Learning}} \bibinfo{volume}{19}, \bibinfo{number}{2}
  (\bibinfo{year}{2004}), \bibinfo{pages}{135--149}.
\newblock


\bibitem[\protect\citeauthoryear{Garvin-Doxas and Barker}{Garvin-Doxas and
  Barker}{2004}]%
        {garvin2004communication}
\bibfield{author}{\bibinfo{person}{Kathy Garvin-Doxas} {and}
  \bibinfo{person}{Lecia~J Barker}.} \bibinfo{year}{2004}\natexlab{}.
\newblock \showarticletitle{Communication in Computer Science Classrooms:
  Understanding Defensive Climates as a Means of Creating Supportive
  Behaviors}.
\newblock \bibinfo{journal}{\emph{Journal on Educational Resources in Computing
  (JERIC)}} \bibinfo{volume}{4}, \bibinfo{number}{1} (\bibinfo{year}{2004}),
  \bibinfo{pages}{2--es}.
\newblock


\bibitem[\protect\citeauthoryear{Gopal, Miranda, Robichaux, and Bostrom}{Gopal
  et~al\mbox{.}}{1997}]%
        {gopal1997leveraging}
\bibfield{author}{\bibinfo{person}{Abhijit Gopal}, \bibinfo{person}{Shaila~M
  Miranda}, \bibinfo{person}{Barry~P Robichaux}, {and}
  \bibinfo{person}{Robert~P Bostrom}.} \bibinfo{year}{1997}\natexlab{}.
\newblock \showarticletitle{{Leveraging Diversity with Information Technology:
  Gender, Attitude, and Intervening Influences in the Use of Group Support
  Systems}}.
\newblock \bibinfo{journal}{\emph{Small Group Research}} \bibinfo{volume}{28},
  \bibinfo{number}{1} (\bibinfo{year}{1997}), \bibinfo{pages}{29--71}.
\newblock


\bibitem[\protect\citeauthoryear{Hill, Song, and West}{Hill
  et~al\mbox{.}}{2009}]%
        {hill2009social}
\bibfield{author}{\bibinfo{person}{Janette~R Hill}, \bibinfo{person}{Liyan
  Song}, {and} \bibinfo{person}{Richard~E West}.}
  \bibinfo{year}{2009}\natexlab{}.
\newblock \showarticletitle{Social Learning Theory and Web-Based Learning
  Environments: A Review of Research and Discussion of Implications}.
\newblock \bibinfo{journal}{\emph{The Amer. Jrnl. of Distance Education}}
  \bibinfo{volume}{23}, \bibinfo{number}{2} (\bibinfo{year}{2009}),
  \bibinfo{pages}{88--103}.
\newblock


\bibitem[\protect\citeauthoryear{Joyner, Bernstein, Bolger, Dittamo, Gorham,
  and Hudson}{Joyner et~al\mbox{.}}{2022}]%
        {joynerDoubleEdged}
\bibfield{author}{\bibinfo{person}{David~A. Joyner}, \bibinfo{person}{Lily
  Bernstein}, \bibinfo{person}{Ian Bolger}, \bibinfo{person}{Maria-Isabelle
  Dittamo}, \bibinfo{person}{Stephanie Gorham}, {and} \bibinfo{person}{Rachel
  Hudson}.} \bibinfo{year}{2022}\natexlab{}.
\newblock \showarticletitle{{Anonymity: A Double-Edged Sword for Gender Equity
  in a CS1 Forum?}}. In \bibinfo{booktitle}{\emph{Proceedings of the 53rd ACM
  Technical Symposium on Computer Science Education V. 1}} (Providence, RI,
  USA) \emph{(\bibinfo{series}{SIGCSE 2022})}. \bibinfo{publisher}{Association
  for Computing Machinery}, \bibinfo{address}{New York, NY, USA},
  \bibinfo{pages}{766–772}.
\newblock
\showISBNx{9781450390705}
\urldef\tempurl%
\url{https://doi.org/10.1145/3478431.3499289}
\showDOI{\tempurl}


\bibitem[\protect\citeauthoryear{Karau and Williams}{Karau and
  Williams}{2014}]%
        {karau2014understanding}
\bibfield{author}{\bibinfo{person}{Steven~J Karau} {and}
  \bibinfo{person}{Kipling~D Williams}.} \bibinfo{year}{2014}\natexlab{}.
\newblock \showarticletitle{Understanding individual motivation in groups: The
  collective effort model}.
\newblock In \bibinfo{booktitle}{\emph{Groups at work}}.
  \bibinfo{publisher}{Psychology Press}, \bibinfo{pages}{127--156}.
\newblock


\bibitem[\protect\citeauthoryear{Kim}{Kim}{2013}]%
        {kim2013influence}
\bibfield{author}{\bibinfo{person}{Jungjoo Kim}.}
  \bibinfo{year}{2013}\natexlab{}.
\newblock \showarticletitle{Influence of group size on students' participation
  in online discussion forums}.
\newblock \bibinfo{journal}{\emph{Computers \& Education}}
  \bibinfo{volume}{62} (\bibinfo{year}{2013}), \bibinfo{pages}{123--129}.
\newblock


\bibitem[\protect\citeauthoryear{Kirkpatrick and Mayfield}{Kirkpatrick and
  Mayfield}{2017}]%
        {kirkpatrick2017evaluating}
\bibfield{author}{\bibinfo{person}{Michael~S Kirkpatrick} {and}
  \bibinfo{person}{Chris Mayfield}.} \bibinfo{year}{2017}\natexlab{}.
\newblock \showarticletitle{Evaluating an Alternative CS1 for Students with
  Prior Programming Experience}. In \bibinfo{booktitle}{\emph{Proceedings of
  the 2017 ACM SIGCSE Technical Symposium on Computer Science Education}}.
  \bibinfo{pages}{333--338}.
\newblock


\bibitem[\protect\citeauthoryear{Kraut and Resnick}{Kraut and Resnick}{2012}]%
        {kraut2012building}
\bibfield{author}{\bibinfo{person}{Robert~E Kraut} {and} \bibinfo{person}{Paul
  Resnick}.} \bibinfo{year}{2012}\natexlab{}.
\newblock \bibinfo{booktitle}{\emph{Building successful online communities:
  Evidence-based social design}}.
\newblock \bibinfo{publisher}{MIT Press}.
\newblock


\bibitem[\protect\citeauthoryear{K{\"u}{\c{c}}{\"u}k}{K{\"u}{\c{c}}{\"u}k}{2010}]%
        {kuccuk2010lurking}
\bibfield{author}{\bibinfo{person}{Mestan K{\"u}{\c{c}}{\"u}k}.}
  \bibinfo{year}{2010}\natexlab{}.
\newblock \showarticletitle{Lurking in online asynchronous discussion}.
\newblock \bibinfo{journal}{\emph{Procedia-Social and Behavioral Sciences}}
  \bibinfo{volume}{2}, \bibinfo{number}{2} (\bibinfo{year}{2010}),
  \bibinfo{pages}{2260--2263}.
\newblock


\bibitem[\protect\citeauthoryear{La~Vista, Falkner, and Szabo}{La~Vista
  et~al\mbox{.}}{2017}]%
        {la2017understanding}
\bibfield{author}{\bibinfo{person}{Daniel La~Vista}, \bibinfo{person}{Nickolas
  Falkner}, {and} \bibinfo{person}{Claudia Szabo}.}
  \bibinfo{year}{2017}\natexlab{}.
\newblock \showarticletitle{Understanding the Effects of Intervention on
  Computer Science Student Behaviour in On-line Forums}. In
  \bibinfo{booktitle}{\emph{Proceedings of the 2017 ACM Conference on
  Innovation and Technology in Computer Science Education}}.
  \bibinfo{pages}{200--205}.
\newblock


\bibitem[\protect\citeauthoryear{Lake}{Lake}{1999}]%
        {lake1999reducing}
\bibfield{author}{\bibinfo{person}{David Lake}.}
  \bibinfo{year}{1999}\natexlab{}.
\newblock \showarticletitle{Reducing Isolation for Distance Students: An
  On‐line Initiative}.
\newblock \bibinfo{journal}{\emph{Open Learning: The Journal of Open, Distance
  and e-Learning}} \bibinfo{volume}{14}, \bibinfo{number}{3}
  (\bibinfo{year}{1999}), \bibinfo{pages}{14--23}.
\newblock


\bibitem[\protect\citeauthoryear{May, Wachs, and Hann{\'a}k}{May
  et~al\mbox{.}}{2019}]%
        {may2019gender}
\bibfield{author}{\bibinfo{person}{Anna May}, \bibinfo{person}{Johannes Wachs},
  {and} \bibinfo{person}{Anik{\'o} Hann{\'a}k}.}
  \bibinfo{year}{2019}\natexlab{}.
\newblock \showarticletitle{{Gender Differences in Participation and Reward on
  Stack Overflow}}.
\newblock \bibinfo{journal}{\emph{Empirical Software Engineering}}
  \bibinfo{volume}{24}, \bibinfo{number}{4} (\bibinfo{year}{2019}),
  \bibinfo{pages}{1997--2019}.
\newblock


\bibitem[\protect\citeauthoryear{Mazuro and Rao}{Mazuro and Rao}{2011}]%
        {mazuro2011online}
\bibfield{author}{\bibinfo{person}{Cath Mazuro} {and} \bibinfo{person}{Namrata
  Rao}.} \bibinfo{year}{2011}\natexlab{}.
\newblock \showarticletitle{Online Discussion Forums in Higher Education: Is
  `Lurking’ Working?}
\newblock \bibinfo{journal}{\emph{International Journal for Cross-Disciplinary
  Subjects in Education}} \bibinfo{volume}{2}, \bibinfo{number}{2}
  (\bibinfo{year}{2011}), \bibinfo{pages}{364--371}.
\newblock


\bibitem[\protect\citeauthoryear{Mendez, Padala, Steine-Hanson, Hilderbrand,
  Horvath, Hill, Simpson, Patil, Sarma, and Burnett}{Mendez
  et~al\mbox{.}}{2018}]%
        {mendez2018open}
\bibfield{author}{\bibinfo{person}{Christopher Mendez},
  \bibinfo{person}{Hema~Susmita Padala}, \bibinfo{person}{Zoe Steine-Hanson},
  \bibinfo{person}{Claudia Hilderbrand}, \bibinfo{person}{Amber Horvath},
  \bibinfo{person}{Charles Hill}, \bibinfo{person}{Logan Simpson},
  \bibinfo{person}{Nupoor Patil}, \bibinfo{person}{Anita Sarma}, {and}
  \bibinfo{person}{Margaret Burnett}.} \bibinfo{year}{2018}\natexlab{}.
\newblock \showarticletitle{{Open source barriers to entry, revisited: A
  sociotechnical perspective}}. In \bibinfo{booktitle}{\emph{Proceedings of the
  40th International conference on software engineering}}.
  \bibinfo{pages}{1004--1015}.
\newblock


\bibitem[\protect\citeauthoryear{Mustafaraj and Bu}{Mustafaraj and Bu}{2015}]%
        {mustafaraj2015visible}
\bibfield{author}{\bibinfo{person}{Eni Mustafaraj} {and}
  \bibinfo{person}{Jessica Bu}.} \bibinfo{year}{2015}\natexlab{}.
\newblock \showarticletitle{The Visible and Invisible in a MOOC Discussion
  Forum}. In \bibinfo{booktitle}{\emph{Proceedings of the Second (2015) ACM
  Conference on Learning@ Scale}}. \bibinfo{pages}{351--354}.
\newblock


\bibitem[\protect\citeauthoryear{Nandi, Hamilton, Harland, and Warburton}{Nandi
  et~al\mbox{.}}{2011}]%
        {nandi2011active}
\bibfield{author}{\bibinfo{person}{Dip Nandi}, \bibinfo{person}{Margaret
  Hamilton}, \bibinfo{person}{James Harland}, {and} \bibinfo{person}{Geoff
  Warburton}.} \bibinfo{year}{2011}\natexlab{}.
\newblock \showarticletitle{How active are students in online discussion
  forums?}. In \bibinfo{booktitle}{\emph{Proceedings of the Thirteenth
  Australasian Computing Education Conference-Volume 114}}.
  \bibinfo{pages}{125--134}.
\newblock


\bibitem[\protect\citeauthoryear{Nonnecke and Preece}{Nonnecke and
  Preece}{2001}]%
        {nonnecke2001lurkers}
\bibfield{author}{\bibinfo{person}{Blair Nonnecke} {and} \bibinfo{person}{Jenny
  Preece}.} \bibinfo{year}{2001}\natexlab{}.
\newblock \showarticletitle{Why lurkers lurk}.
\newblock  (\bibinfo{year}{2001}).
\newblock


\bibitem[\protect\citeauthoryear{Palmer, Holt, and Bray}{Palmer
  et~al\mbox{.}}{2008}]%
        {palmer2008does}
\bibfield{author}{\bibinfo{person}{Stuart Palmer}, \bibinfo{person}{Dale Holt},
  {and} \bibinfo{person}{Sharyn Bray}.} \bibinfo{year}{2008}\natexlab{}.
\newblock \showarticletitle{Does the discussion help? The impact of a formally
  assessed online discussion on final student results}.
\newblock \bibinfo{journal}{\emph{British Journal of Educational Technology}}
  \bibinfo{volume}{39}, \bibinfo{number}{5} (\bibinfo{year}{2008}),
  \bibinfo{pages}{847--858}.
\newblock


\bibitem[\protect\citeauthoryear{Ruthotto, Kreth, Stevens, Trively, and
  Melkers}{Ruthotto et~al\mbox{.}}{2020}]%
        {ruthotto2020lurking}
\bibfield{author}{\bibinfo{person}{Isabel Ruthotto}, \bibinfo{person}{Quintin
  Kreth}, \bibinfo{person}{Jillian Stevens}, \bibinfo{person}{Clare Trively},
  {and} \bibinfo{person}{Julia Melkers}.} \bibinfo{year}{2020}\natexlab{}.
\newblock \showarticletitle{Lurking and participation in the virtual classroom:
  The effects of gender, race, and age among graduate students in computer
  science}.
\newblock \bibinfo{journal}{\emph{Computers \& Education}}
  \bibinfo{volume}{151} (\bibinfo{year}{2020}), \bibinfo{pages}{103854}.
\newblock


\bibitem[\protect\citeauthoryear{Ryle and Cumming}{Ryle and Cumming}{2007}]%
        {ryle2007reflections}
\bibfield{author}{\bibinfo{person}{Anita Ryle} {and} \bibinfo{person}{Kaye
  Cumming}.} \bibinfo{year}{2007}\natexlab{}.
\newblock \showarticletitle{Reflections on Engagement in Online Learning
  Communities}.
\newblock \bibinfo{journal}{\emph{International Journal of Pedagogies and
  Learning}} \bibinfo{volume}{3}, \bibinfo{number}{3} (\bibinfo{year}{2007}),
  \bibinfo{pages}{35--46}.
\newblock


\bibitem[\protect\citeauthoryear{Shaw}{Shaw}{2013}]%
        {shaw2013relationships}
\bibfield{author}{\bibinfo{person}{Ruey-Shiang Shaw}.}
  \bibinfo{year}{2013}\natexlab{}.
\newblock \showarticletitle{The relationships among group size, participation,
  and performance of programming language learning supported with online
  forums}.
\newblock \bibinfo{journal}{\emph{Computers \& Education}}
  \bibinfo{volume}{62} (\bibinfo{year}{2013}), \bibinfo{pages}{196--207}.
\newblock


\bibitem[\protect\citeauthoryear{Sobel, Gilmartin, and Sankar}{Sobel
  et~al\mbox{.}}{2016}]%
        {sobel2016class}
\bibfield{author}{\bibinfo{person}{Melissa Sobel}, \bibinfo{person}{Jessica
  Gilmartin}, {and} \bibinfo{person}{Pooja Sankar}.}
  \bibinfo{year}{2016}\natexlab{}.
\newblock \showarticletitle{{Class Size and Confidence Levels Among Female STEM
  Students} [Impact]}.
\newblock \bibinfo{journal}{\emph{IEEE Technology and Society Magazine}}
  \bibinfo{volume}{35}, \bibinfo{number}{1} (\bibinfo{year}{2016}),
  \bibinfo{pages}{23--26}.
\newblock


\bibitem[\protect\citeauthoryear{Sun, Rau, and Ma}{Sun et~al\mbox{.}}{2014}]%
        {sun2014understanding}
\bibfield{author}{\bibinfo{person}{Na Sun}, \bibinfo{person}{Patrick Pei-Luen
  Rau}, {and} \bibinfo{person}{Liang Ma}.} \bibinfo{year}{2014}\natexlab{}.
\newblock \showarticletitle{Understanding lurkers in online communities: A
  literature review}.
\newblock \bibinfo{journal}{\emph{Computers in Human Behavior}}
  \bibinfo{volume}{38} (\bibinfo{year}{2014}), \bibinfo{pages}{110--117}.
\newblock


\bibitem[\protect\citeauthoryear{Tafliovich, Campbell, and Petersen}{Tafliovich
  et~al\mbox{.}}{2013}]%
        {tafliovich2013student}
\bibfield{author}{\bibinfo{person}{Anya Tafliovich}, \bibinfo{person}{Jennifer
  Campbell}, {and} \bibinfo{person}{Andrew Petersen}.}
  \bibinfo{year}{2013}\natexlab{}.
\newblock \showarticletitle{A student perspective on prior experience in CS1}.
  In \bibinfo{booktitle}{\emph{Proceeding of the 44th ACM Technical Symposium
  on Computer Science Education}}. \bibinfo{pages}{239--244}.
\newblock


\bibitem[\protect\citeauthoryear{Terrell, Kofink, Middleton, Rainear,
  Murphy-Hill, Parnin, and Stallings}{Terrell et~al\mbox{.}}{2017}]%
        {terrell2017gender}
\bibfield{author}{\bibinfo{person}{Josh Terrell}, \bibinfo{person}{Andrew
  Kofink}, \bibinfo{person}{Justin Middleton}, \bibinfo{person}{Clarissa
  Rainear}, \bibinfo{person}{Emerson Murphy-Hill}, \bibinfo{person}{Chris
  Parnin}, {and} \bibinfo{person}{Jon Stallings}.}
  \bibinfo{year}{2017}\natexlab{}.
\newblock \showarticletitle{{Gender differences and bias in open source: Pull
  request acceptance of women versus men}}.
\newblock \bibinfo{journal}{\emph{PeerJ Computer Science}}  \bibinfo{volume}{3}
  (\bibinfo{year}{2017}), \bibinfo{pages}{e111}.
\newblock


\bibitem[\protect\citeauthoryear{Thinnyun, Lenfant, Pettit, and Hott}{Thinnyun
  et~al\mbox{.}}{2021}]%
        {thinnyun2021gender}
\bibfield{author}{\bibinfo{person}{Adrian Thinnyun}, \bibinfo{person}{Ryan
  Lenfant}, \bibinfo{person}{Raymond Pettit}, {and} \bibinfo{person}{John~R
  Hott}.} \bibinfo{year}{2021}\natexlab{}.
\newblock \showarticletitle{{Gender and Engagement in CS Courses on Piazza}}.
  In \bibinfo{booktitle}{\emph{Proceedings of the 52nd ACM Technical Symposium
  on Computer Science Education}}. \bibinfo{pages}{438--444}.
\newblock


\bibitem[\protect\citeauthoryear{Weller}{Weller}{2000}]%
        {weller2000implementing}
\bibfield{author}{\bibinfo{person}{Martin Weller}.}
  \bibinfo{year}{2000}\natexlab{}.
\newblock \showarticletitle{Implementing a CMC tutor group for an existing
  distance education course}.
\newblock \bibinfo{journal}{\emph{Journal of Computer Assisted Learning}}
  \bibinfo{volume}{16}, \bibinfo{number}{3} (\bibinfo{year}{2000}),
  \bibinfo{pages}{178--183}.
\newblock


\bibitem[\protect\citeauthoryear{Wong-Aitken, Cukierman, and
  Chilana}{Wong-Aitken et~al\mbox{.}}{2022}]%
        {wong2022depends}
\bibfield{author}{\bibinfo{person}{David Wong-Aitken}, \bibinfo{person}{Diana
  Cukierman}, {and} \bibinfo{person}{Parmit~K Chilana}.}
  \bibinfo{year}{2022}\natexlab{}.
\newblock \showarticletitle{“It Depends on Whether or Not I’m Lucky”: How
  Students in an Introductory Programming Course Discover, Select, and Assess
  the Utility of Web-Based Resources}.
\newblock  (\bibinfo{year}{2022}).
\newblock


\bibitem[\protect\citeauthoryear{Ying, Rodr{\'\i}guez, Dibble, Martin, Boyer,
  Thomas, and Gilbert}{Ying et~al\mbox{.}}{2021}]%
        {ying2021confidence}
\bibfield{author}{\bibinfo{person}{Kimberly~Michelle Ying},
  \bibinfo{person}{Fernando~J Rodr{\'\i}guez},
  \bibinfo{person}{Alexandra~Lauren Dibble}, \bibinfo{person}{Alexia~Charis
  Martin}, \bibinfo{person}{Kristy~Elizabeth Boyer},
  \bibinfo{person}{Sanethia~V Thomas}, {and} \bibinfo{person}{Juan~E Gilbert}.}
  \bibinfo{year}{2021}\natexlab{}.
\newblock \showarticletitle{{Confidence, Connection, and Comfort: Reports from
  an All-Women's CS1 Class}}. In \bibinfo{booktitle}{\emph{Proceedings of the
  52nd ACM Technical Symposium on Computer Science Education}}.
  \bibinfo{pages}{699--705}.
\newblock


\end{thebibliography}
\end{document}